\documentclass[12pt]{article}
\usepackage{epsfig}
\usepackage{axodraw}
\def\be{\begin{equation}}
\def\ee{\end{equation}}
\def\beq{\begin{equation}}
\def\eeq{\end{equation}}
\def\beqar{\begin{eqnarray}}
\def\eeqar{\end{eqnarray}}
\def\barr{\begin{array}}
\def\earr{\end{array}}

\def\and{\qquad {\rm and } \qquad}

\def\slp{p \hspace{-1ex}/}

\def\slp{p \hspace{-1ex}/}
\def\sls{s \hspace{-1ex}/}

\def\sp{\vec{s}_+}
\def\sm{\vec{s}_-}
\def\SP{(\vec{s}_+ +\vec{s}_-)}
\def\SM{(\vec{s}_+-\vec{s}_-)}
\def\HP{(h_-\vec{s}_+ + h_+\vec{s}_-)}
\def\HM{(h_-\vec{s}_+ - h_+\vec{s}_-)}
\def\hp{h_+}
\def\hm{h_-}
\def\Kv{\vec{K}}
\def\pv{\vec{p}}
\def\pvp{\vec{p}_+}
\def\rv{\vec{r}}

\def\rvt{\vec{r}^T}
\def\uvt{\vec{u}^T}

\def\tvt{\vec{t}^T}
\def\uv{\vec{u}}

\def\tv{\vec{t}}
\def\nv{\vec{n}}
\def\g5{\gamma_5}
\newcommand{\nc}{\newcommand}
\nc{\jc}{\frac{1}{4}}  \nc{\sll}{S_{LL}}     \nc{\slr}{S_{LR}}
\nc{\srl}{S_{RL}}      \nc{\srr}{S_{RR}}     \nc{\vll}{V_{LL}}
\nc{\vlr}{V_{LR}}      \nc{\vrl}{V_{RL}}     \nc{\vrr}{V_{RR}}
\nc{\tll}{T_{LL}}      \nc{\tlrs}{T_{LR}}    \nc{\trl}{T_{RL}}
\nc{\trr}{T_{RR}}      \nc{\slld}{S_{LL}^D}  \nc{\slrd}{S_{LR}^D}
\nc{\srld}{S_{RL}^D}   \nc{\srrd}{S_{RR}^D}  \nc{\vlld}{V_{LL}^D}
\nc{\vlrd}{V_{LR}^D}   \nc{\vrld}{V_{RL}^D}  \nc{\vrrd}{V_{RR}^D}
\nc{\tlld}{T_{LL}^D}   \nc{\tlrd}{T_{LR}^D}  \nc{\trld}{T_{RL}^D}
\nc{\trrd}{T_{RR}^D}   \nc{\aqde}{\alpha_{qde}}
\nc{\alq}{\alpha_{\ell q}}        \nc{\alqp}{\alpha_{\ell q'}}
\nc{\alqt}{\alpha_{\ell q}^{(3)}} \nc{\alqtc}{\alpha_{\ell
q}^{(3)*}} \nc{\alqj}{\alpha_{\ell q}^{(1)}}
\nc{\alqjc}{\alpha_{\ell q}^{(1)*}} \nc{\aeu}{\alpha_{eu}}
\nc{\alu}{\alpha_{\ell u}} \nc{\aqe}{\alpha_{qe}}
\nc{\ber}{\begin{eqnarray*}} \nc{\enr}{\end{eqnarray*}}
\nc{\jmpb}{(1-\beta)/(1+\beta)} \nc{\wspR}{r}      \nc{\varx}{x}
\nc{\bt}{\beta}

\nc{\non}{\nonumber} \nc{\lspace}{\;\;\;\;\;\;\;\;\;\;}
\nc{\llspace}{\lspace \lspace}
\nc{\jnl}{\frac{1}{{\mit\Lambda}^2}} \nc{\jd}{\frac{1}{2}}
\nc{\comment}[1]{}
\begin{document}
%\thispagestyle{empty}
%\setcounter{page}{0}
%\renewcommand{\thefootnote}{\fnsymbol{footnote}}
%\begin{flushright}
%hep-ph/yymmnnn
%\end{flushright}
\vskip .3cm

\begin{center}{\Large \bf \boldmath
Two-particle kinematic distributions from new physics at an
electron-positron collider with polarized beams}
\vskip 1cm
{B. Ananthanarayan$^a$, Saurabh D. Rindani$^{b}$}
\vskip .5cm

{\it $^a$Centre for High Energy Physics,
Indian Institute of Science\\ Bangalore
560 012, India\\~ \\
$^b$Theoretical Physics Division, Physical Research Laboratory\\
Navrangpura, Ahmedabad 380 009,
India}
\end{center}

\begin{quote}
\begin{abstract}
The kinematic distributions in two-particle inclusive processes 
at an $e^+e^-$ collider arising from standard-model 
$s$-channel exchange of a virtual $\gamma$ or $Z$ 
and the interference of the standard-model
 contribution with contributions from physics
beyond the standard model involving $s$-channel exchanges are derived 
entirely in terms
of the space-time signature of such new physics.  
Transverse as well as longitudinal polarizations of the electron and
positron beams are taken into account.
We show how these model-independent distributions can be used to 
deduce some general properties of the
nature of the interaction.
We then specialize to two specific two-particle final states, 
viz., $ZH$, where $H$ is one of the Higgs bosons in a model with an
extended Higgs sector, and $f\bar f$, where $f$, $\bar f$ are a pair of
conjugate charged fermions, wherein distributions
of two (of the possibly several) decay products are measured.
We show how some of the properties of the distributions have been 
realized in the analysis of physics beyond the standard model in earlier
work which made use of two-particle angular distributions. 
\end{abstract}
\end{quote}
\newpage
\section{Introduction}\label{intro}

The proposed International Linear Collider (ILC) which could 
collide $e^+$ and $e^-$
at a centre of mass energy of several hundred GeV, if built, would serve
as an instrument for precision measurements of various parameters
underlying particle physics~\cite{LC_SOU}.
A strong beam polarization programme of transverse or
longitudinal beam polarization is also being seriously proposed by
investigators in the field~\cite{POWER}.
Besides carrying out precision studies
of properties of standard-model (SM) particles, the ILC is geared to probe
physics beyond the standard model (BSM).  In particular, the ILC will be
sensitive to BSM physics even if the energy is not sufficient
to produce BSM particles directly, via precision studies that
are sensitive to the propagation of such particles in loops.  
BSM physics can manifest itself in a variety of new effects including
CP violation, momentum correlations of SM particles, correlations
involving
spins of decaying particles as well as spins of the electron and positron
beams.  

Recently, we presented an approach that relies on the
characterization of new physics in terms of its space-time
transformation properties~\cite{BASDR} using one-particle inclusive
distributions in $e^+e^-$ annihilation. 
This approach was model independent, and relied only on
the use of Lorentz covariance for deriving the most general form of
one-particle kinematic distributions for the cases where the BSM
interaction had different space-time properties. The distributions were
expressed in terms of Lorentz-invariant `structure functions' much as
kinematic distributions in deep-inelastic scattering are characterized
in terms of structure functions in what is now standard treatment. 
We demonstrated the
utility of such an approach for a general analysis of different types of
processes. 

For processes where the single particle of interest is
heavy, it is unstable and is therefore 
invariably detected by means of its decay products. Thus, the process 
then has at least two particles in the final state whose momenta are
measured, and an extension to two-particle distributions would be
useful. In addition to this motivation, a two-particle inclusive
distribution would carry more information than a one-particle one.
In fact, there can be qualitatively new
information in the two-particle case. For instance, as we will see later, 
in certain situations, signatures of CP violation are absent in the
one-particle exclusive final case, but appear naturally in the
two-particle case. 
Keeping these motivations in mind,
we have now extended our study to the two-particle inclusive process 
$e^+ e^-\to h_1(p_1) h_2(p_2) X$, where $h_1$ and $h_2$ denote two SM
particles that are detected, and $p_1$ and $p_2$ are their
respective momenta.  The latter process is depicted in Fig.~\ref{basicfig}.
It may be noted that this general process actually encompasses a class of 
different exclusive
processes, including those where $h_1$ and $h_2$ arise from the decay of
a heavy particle or a resonance. 

In spirit it is the extension of
the pioneering work of Dass and Ross~\cite{DR1,DR2} that had
been performed in the context of $\gamma$ contributing
to the $s$-channel production, probing the then undiscovered neutral
current.  Our work in practice is the inclusion of 
$Z$ in the s-channel,
in addition to $\gamma$, and 
 where now it is  the as-yet undiscovered new physics that we intend to
probe.  Significant new features arise
due to the presence of the axial-vector coupling of the $Z$
to the electron, a feature missing in a vector theory like QED.
Indeed, a significant feature is that here too there are additional
structure functions compared to the analysis presented
in ref.~\cite{BASDR}.
These are absent in the case of the reaction where only one
particle was detected with no spin information.
It may also be emphasized that once a general discussion is provided
for an inclusive final state, it may be readily applied to
exclusive final states as well, thereby providing a framework
for studying several processes of interest.
More importantly, the inclusive state could simply arise from the decays
of the particles in a two-particle final state.

\begin{center}
\begin{figure}[htb] \label{basicfig}
\begin{picture}(500,100)(-50,0)
\ArrowLine(10,100)(41.5,60.5)
\Text(-10,90)[]{$e^+(p_+)$}
\Text(85,75)[]{$h_1(p_1)$}
\Text(85,40)[]{$h_2(p_2)$}
\Text(85,60)[]{$X$}
\Text(135,60)[]{$=$}
\Text(300,60)[]{$+$}
{\GOval(45,60)(5,5)(0){0.5}}
\ArrowLine(10,20)(39.5,59.5)
\Text(-10,40)[]{$e^-(p_-)$}
\ArrowLine(49,61)(80,100)
\ArrowLine(49,59)(80,20)
\ArrowLine(50,60)(80,60)
%%%%%%%%%%%%%%%%%%%%
\ArrowLine(160,100)(190,60)
\ArrowLine(160,20)(190,60)
\Photon(190,60)(250,60){2}{6}
\Text(220,80)[]{$\gamma, \, Z$}
\ArrowLine(250,60)(280,100)
\ArrowLine(250,60)(280,20)
\ArrowLine(250,60)(280,60)
%%%%%%%%%%%%%%%%%%%%%%%%%%%%%%%
\ArrowLine(310,100)(340,60)
\ArrowLine(310,20)(340,60)
\ArrowLine(342,60)(372,100)
\ArrowLine(342,60)(372,20)
\ArrowLine(342,60)(372,60)
{\GOval(341,60)(2,2)(0){0}}
\Text(343,45)[]{\small NP}
\end{picture}
\caption{The basic process}
\end{figure}
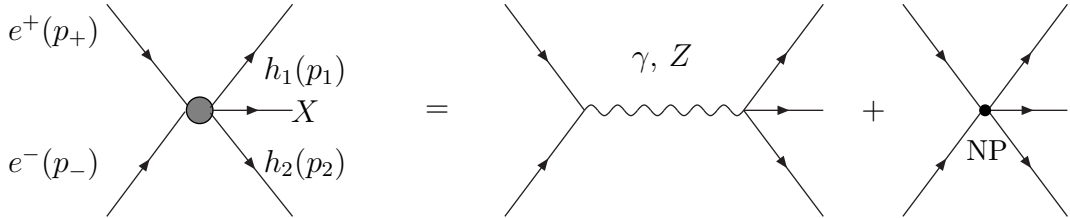
\end{center}

Our formalism is restricted to the broad framework originally utilized
in \cite{DR1,DR2}, which envisages new physics only through an
$s$-channel exchange. Thus the BSM effects could arise through the
exchange of a new particle like a new gauge boson $Z'$, or through the
exchange of $Z$, but a with a BSM vertex or a SM loop producing the 
final state in question. Moreover, the SM contribution is also assumed
to be through the tree-level exchange of a virtual photon and a virtual
$Z$. There are other interesting scenarios where the present formalism
is not applicable without modification. For example, in the production
of a SM gauge-boson pair in the final state, the SM contribution is via
a $t$-channel exchange of an electron, not amenable to description in
the present formalism. However, suitable modifications of our approach,
to be pursued in future, suggest themselves to deal with such
situations. A modified formalism for the $\gamma Z$ final
state was discussed in \cite{BASDR}, and the results compared with those
in \cite{BASDRgammaZ,basdrcont}.

The expectations from our general model-independent analysis is shown for 
some specific processes to be consistent with the results obtained
earlier for those processes. Our approach would thus be useful to derive
general results for newer processes which fall within the framework
described above. We would thus be able to anticipate certain results
without a detailed calculation for each individual process.

In Sec.~\ref{correlations} we present a computation of the spin-momentum
correlations resulting from the presence of structure functions
that characterize the new physics.  Our results here are presented
in the form of results arising from the computation of a trace that
encodes the leptonic tensor as well as the new physics encoded in
a tensor constructed out of the momenta of the observed final-state
particles (what is known as a `hadronic' tensor, for historical reasons,
since the term arose at a time 
when the final state consisted largely of hadrons).  
These tables provide the
analogue for the SM and new physics, of what was 
provided by Dass and Ross~\cite{DR2}
for QED and neutral currents.  In Sec.~\ref{characterization} 
we discuss the CP and T properties of correlations for different classes
of inclusive and exclusive final states. In Sec.~\ref{sec:compol}
we provide a discussion on the the polarization dependence of the
correlations in different cases.
In Sec.~\ref{exclusive} we will specialize to two specific examples of 
processes, into which our approach 
 can give significant insight.  
In Sec.~\ref{conclusions} we present our conclusions and discuss
prospects for extension of the present framework to account for classes
of BSM interactions not presently covered.

\section{Computation of correlations}\label{correlations}

We consider the two-particle inclusive process
\begin{eqnarray}\label{process}
&
e^-(p_-) + e^+(p_+) \to h_1(p_1) + h_2(p_2)+ X, & 
%e^-(p_-) + e^+(p_+) \to h(p,s)             X, & \nonumber \\
\end{eqnarray}
where $h_1$ and $h_2$ are final state particles whose momenta are measured,
but not their spin, and
$X$ is an inclusive state.
The process is assumed to occur through an $s$-channel
exchange of a photon and a $Z$ in the SM, and through a  new current whose
coupling to $e^+e^-$ can be of the type $V,A$, or $S,P$, or $T$.

Since we will deal with a general case without specifying the nature or
couplings of $h_1, h_2$, 
we do not attempt to write the amplitude for the process
(\ref{process}). We will only obtain the general form, for each case of
the new coupling, of the contribution to the angular distribution of
$h_1, h_2$ 
from the interference of the SM amplitude with the new physics
amplitude.

It might be clarified here that even though we use the term
``inclusive'' implying that no measurement is made on the state $X$, in
practice it may be that the state $X$ is restricted to a concrete one-particle
or two-particle state which is detected. In such a case the sum is not
over all possible states $X$. Nevertheless, the momenta of the few
particles in the state $X$ are assumed to be integrated over, so that
there is a gain in statistics as compared to a completely exclusive
measurement. The angular distributions we calculate hold also for such a
case, except that structure functions would depend on the states
included in $X$.

Following Dass and Ross \cite{DR1,DR2}, we calculate the relevant factor in
the interference between the standard model currents with the
BSM currents as
\begin{equation}\label{trace}
{\rm Tr}[(1-\g5 \hp + \g5 \sls_+)\slp_+\gamma_\mu(g_V^e-g_A^e \gamma_5)
(1+\g5 h_-+\g5 \sls_-)\slp_-\Gamma_i]H^{i\mu }.
\end{equation}
Here $g_V^e, g_A^e$ are the vector and axial-vector couplings of the
photon or $Z$ to the electron current, and $\Gamma_i$ is the
corresponding coupling to the new-physics current, 
$h_{\pm}$ are the
helicities (in units of $\frac{1}{2}$)
of $e^{\pm}$, and $s_{\pm}$ are respectively their transverse polarizations.
For ease of comparison, we have sought to stay with the notation
of refs.~\cite{DR1,DR2}, with some exceptions which we
spell out when necessary.  We should of course add the
contributions coming from photon exchange and $Z$ exchange, with the
appropriate propagator factors. However, we give here the results for
$Z$ exchange, from which the case of photon exchange
 can be deduced as a special
case. The tensor $H^{i\mu }$ stands for the interference between the
couplings of the final state to the SM current and the new-physics
current, summed over final-state polarizations, and over the phase space
of the unobserved particles $X$. It is only a function of the the
momenta $q=p_-+p_+$, $p_1$ and $p_2$. The implied summation over $i$
corresponds to a sum over the forms $V, A, S, P, T$, together with any
Lorentz indices that these may entail.

We now determine the forms of the matrices $\Gamma_i$ and the
tensors $H^{i\mu }$ in the various
cases, using only Lorentz covariance properties.
Our additional currents are as in refs.~\cite{DR1,DR2},
except for the sign of $g_A$ in the following.  We explicitly
note that in our convention is $\epsilon^{0123}=+1$\footnote
{It may be noted that the convention actually used in ref.~\cite{BASDR}
was $\epsilon^{0123}=+1$, with which all the results presented
there being self-consistent.  The corresponding remark made there
about the convention in ref.~\cite{DR1,DR2} may be disregarded.}.
We set the electron mass to zero.  
Consider now the three cases:

\noindent\underline {1. Scalar and pseudoscalar case}:
\smallskip

In this case, there is
no free Lorentz index for the leptonic coupling.
Consequently, we can write it as
\begin{equation}
\Gamma = g_S + i g_P \gamma_5.
\end{equation}
The tensor $H^{i\mu }$ for this case has only one index, viz., $\mu$.
Hence the most general form for $H$ is
\begin{equation}\label{scalarH}
H^{S}_\mu  = (r_\mu - q_\mu {r\cdot q \over q^2}) F^r,
\end{equation}
where $r$ is $p_1$, $p_2$ or $n$ ($n_\rho\equiv \epsilon_{\rho \alpha\beta\gamma}
p_1^\alpha p_2^\beta q^\gamma$).

\bigskip

\noindent\underline {2. Vector and axial-vector case}:
\smallskip

The leptonic
coupling for this case can be written as
\begin{equation}
\Gamma_\nu = \gamma_\nu (g_V -  g_A \gamma_5).
\end{equation}
Note that we differ from Dass and Ross \cite{DR1,DR2} in the sign of the
$g_A$ term.
The tensor $H$ for this case has two indices, and can be written as
\begin{eqnarray}
&
H^V_{\mu\nu} =  -g_{\mu\nu} W_1%(q^2,p\cdot q)
+ {1\over 2}(r_\mu t_\nu+r_\nu t_\mu) W_2^{rt}%(q^2,p\cdot q)
& \nonumber \\
& + \epsilon_{\mu\nu\alpha\beta}u^{\alpha}v^\beta W_3^{uv}%(q^2,p\cdot q),
+{1\over 2}(p_{1\mu} p_{2\nu} - p_{1\nu} p_{2\mu}) W_4, &
\end{eqnarray}
where $W_1, W_2^{rt}, W_3^{uv}, W_4$ are invariant functions,
and $r$, $t$ can be chosen from $p_1$, $p_2$ and $n$, 
and $u,v$ can be chosen from
$p_1$, $p_2$ and $q$. As compared to the one-particle exclusive case,
there is an additional tensor structure with structure function $W_4$,
which requires two particles, being antisymmetric in $p_1$ and $p_2$.

\bigskip

\noindent\underline {3. Tensor case}:
\smallskip

In the tensor case,
the leptonic coupling is
\begin{equation}\label{tensorG}
\Gamma_{\rho\tau} = g_T \sigma_{\rho\tau}.
\end{equation}
The tensor $H$ for this case can be written in terms of the four
invariant functions $F_1$, $F_2$, $PF_1$, $PF_2$ as
\begin{equation}
\begin{array}{lcl}\label{tensorH}
H^T_{\mu\rho\tau}& = & (r_\rho  u_\tau - r_\tau u_\rho ) t_\mu F_1^{rut}
+ ( g_{\rho\mu} r_\tau - g_{\tau\mu} r_\rho ) F_2^r
\\&& + \epsilon_{\rho\tau\alpha\beta} r^\alpha u^\beta t_\mu
PF_1^{rut}
+ \epsilon_{\rho\tau\mu\alpha} r^\alpha PF_2^r,
\end{array}
\end{equation}
where $t$ is chosen from $p_1$, $p_2$ and $n$, 
$u$ from $p_1$, $p_2$, $q$ and $n$, 
$r$ from $p_1$, $p_2$ and $q$. These choices of vectors for $r$, $t$, and
$u$ give a complete set of independent tensors. The use of vectors other
than covered by the choices would result in tensors which are
combinations of tensors described by eq. (\ref{tensorH}). Details can be
found in \cite{DR2}.

The structure functions introduced in the above are functions of the
Lorentz invariants that can be formed from the momenta $q$, $p_1$ and
$p_2$. The dependence of these functions on the Lorentz invariants
encode the dynamics of the BSM interactions. In particular, they would
contain propagators and form factors occurring in the BSM
amplitudes.

We next substitute the leptonic vertices $\Gamma$ and the respective
tensors $H^i$ in  (\ref{trace}), and
evaluate the trace in each case. We present the results in Tables
1-3, with
$\vec K\equiv (\vec{p}_- - \vec{p}_+)/2= E \hat{z}$,
where $\hat{z}$ is a unit vector
in the z-direction, $E$ is the beam energy,
and $\vec{s}_\pm$ lie in the x-y plane.
A superscript $T$ on a vector is used to denote its component transverse
with respect to the $e^+e^-$ beam directions. For example, $\rvt = \rv -
\rv\cdot\hat{z}\,\hat{z}$, and similarly for other vectors.
The tables include, in addition
to results presented in ref.~\cite{DR1,DR2}
with only the $g_V^e$ coupling relevant for QED, also those with $g_A^e$
couplings, relevant for $Z$ exchange in SM.
Tables 1, 2 and 3 are respectively for cases of scalar-pseudoscalar,
vector-axial-vector and tensor couplings respectively.

\begin{table}\label{ps_gVe_table}
\begin{center}
\begin{tabular}{||c|c||}\hline
Term & Correlation                        \\ \hline \hline
${\rm Im}\, (g_P F^r)$ & $2 E^2 \,\rv\cdot [-g_{V}^e\,(\sp-\sm) +g_{A}^e\,(\hp
\sm+\hm \sp)]$            \\
${\rm Im}\, (g_S F^r)$ & $2 E \,\Kv \cdot  [g_{V}^e\, (\sp+\sm)+
g_{A}^e \, (\hp\sm-\hm\sp)]\times \rv$ \\
${\rm Re}\, (g_S F^r)$ & $2 E^2 \,\rv\cdot [ g_{A}^e \, (\sp +  \sm)
+         g_{V}^e \,(\hp \sm - \hm \sp) ]$             \\
${\rm Re}\, (g_P F^r)$ & $2 E \,\Kv \cdot  [ g_{A}^e \,  (\sp -  \sm)
- g_{V}^e\, (\hp \sm + \hm \sp) ]\times \rv$ \\ \hline
\end{tabular}
\caption{List of $S$ and $P$ correlations}
\end{center}
\end{table}

\begin{table}\label{VA_gVe_table}
\begin{center}
\begin{tabular}{||c|c||}\hline
Term & Correlation                       \\ \hline \hline
${\rm Re} \,( g_V W_1)$ & $4 E^2\, [ g_A^e\, (\hp - \hm )
-g_V^e\, (\hp \hm -1) ] $             \\
${\rm Re} \,( g_A W_1)$ & $4 E^2\,[ g_V^e\, (\hp - \hm ) 
-g_A^e\,(\hp \hm -1)] $             \\
${\rm Re} \,( g_V W_2^{rt})$ & $2 E^2\,\left\{
g_A^e\, (\hp-\hm) \rvt\cdot \tvt
-g_V^e\,\left[
 \rvt\cdot \tvt (\hp\hm-1-\sp\cdot \sm)\right.\right. $\\
&$\left.\left. + (\rv\cdot\sm)(\tv\cdot\sp)+
(\rv\cdot\sp)(\tv\cdot\sm)\right]
\right\}$ \\
${\rm Re} \, (g_A W_2^{rt})$ & $ 2 E^2\,\left\{g_V^e\, (\hp-\hm)\rvt\cdot \tvt
-g_A^e\,\left[
 \rvt\cdot \tvt (\hp\hm-1+\sp\cdot \sm) \right.\right. $\\
& $\left. \left. - (\rv\cdot\sm)(\tv\cdot\sp)-
(\rv\cdot\sp)(\tv\cdot\sm)\right]\right\} $             \\
${\rm Im} \,(g_A W_2^{rt})$ & $E\,g_V^e\,[(
\sm\cdot \tv) \rv \cdot (\Kv \times \sp)+
(\sp\cdot \tv) \rv \cdot (\Kv \times \sm)$  \\
&$\;\;\;\;\;\;\; +
(\sm\cdot \rv) \tv \cdot (\Kv \times \sp)+
(\sp\cdot \rv) \tv \cdot (\Kv \times \sm)]$ \\
${\rm Im} \,(g_V W_2^{rt})$ & $\!\!\!\!\! -E\,g_A^e\,[(
\sm\cdot \tv) \rv \cdot (\Kv \times \sp)+
(\sp\cdot \tv) \rv \cdot (\Kv \times \sm)$  \\
&$\;\;\;\;\;\;\; +
(\sm\cdot \rv) \tv \cdot (\Kv \times \sp)+
(\sp\cdot \rv) \tv \cdot (\Kv \times \sm)]$ \\
${\rm Im}\, (g_V W_3^{uv})$ & $4 E^2 (-v^0 u^3+v^3 u^0)\,[-g_V^e\,(\hp-\hm)
+ g_A^e\, (\hp\hm-1)]$ \\
${\rm Im}\, (g_A W_3^{uv})$ & $4 E^2(-v^0 u^3+v^3 u^0)
                               [ -g_A^e\, (\hp- \hm) +g_V^e\,  (\hp \hm-1) ]$             \\
${\rm Im}\, (g_V W_4)$ & $ 2 E (\pv_1 \times \pv_2)\cdot \pvp\,[g_V^e\,(\hp-\hm)
 -g_A^e\, (\hp\hm-1)]$ \\
${\rm Im}\, (g_A W_4)$ & $ 2 E (\pv_1 \times \pv_2)\cdot \pvp\,[
g_A^e\,(\hp- \hm) -g_V^e\,(\hp\hm-1) ]$ \\
\hline
\end{tabular}
\caption{List of $V$ and $A$ correlations }
\end{center}
\end{table}

In the case of $S$, $P$ and $T$ couplings, 
all the entries in the corresponding tables vanish for unpolarized
beams, or for longitudinally polarized beams. 
This is because there is no interference between the SM contribution,
where the coupling to $e^+e^-$ is of the $V$, $A$ (chirality-conserving) type, 
and the contributions with $S$, $P$ or $T$ couplings, which are
chirality violating. Thus at least one beam has to be transversely
polarized to see the interference. Also, in these cases,
it is sufficient to have either
$e^-$ or $e^+$ beams transversely polarized -- it is not necessary for
both beams to have transverse polarization. However, to observe terms
which correspond to combinations like $(h_-\vec s_+ \pm h_+\vec s_-)$, it is
necessary to have at least one beam longitudinally polarized, and the
other transversely polarized. 
Considering that such a configuration, though feasible, is not a simple
option from the experimental point of view, in the practical case when
one beam or both beams are transversely polarized, it can be seen from
Tables 1 and 3 that it is only the coupling $g_V^e$ which goes with the
imaginary part of the structure functions in case of $S$, $P$ couplings,
and $g_A^e$ in case of $T$ couplings. Likewise, $g_A^e$ and $g_V^e$
occur with the real parts in these respective cases.

In case of $V$ and $A$ couplings, 
both beams have to be
polarized, or the effect of polarization vanishes.
It is interesting to note that all the correlations in the latter case are
symmetric under the interchange of $\vec s_+$ and $\vec s_-$.

\begin{table}\label{T_gVe_table}
\begin{center}
\begin{tabular}{||c|c||}\hline
Term & Correlation                         \\ \hline \hline
${\rm Im} (g_T F_1^{rut})$ & $4 E^2
\{  [(\rvt\cdot\tvt)\uv - (\uvt\cdot \tvt)\rv]
\cdot [ g_A^e\, \SM -g_V^e\, \HP ]  $ \\
& $+ (r^0 u^3-r^3 u^0) \tv \cdot [ g_A^e\, \SP -g_V^e \,\HM ] \} $ \\
${\rm Im} (g_T F_2^r)$ &  $4 E^2 \rv \cdot [ - g_A^e\, \SM +g_V^e\,\HP ]$ \\
${\rm Im} (g_T PF_1^{rut}) $ &
$ 4 E\{E\, [ r^0 (\uv \times \tvt) - u^0(\rv\times \tvt)
        ]\cdot [ - g_A^e\, \SM +g_V^e\, \HP ]$  \\
& $-\{[(\rv\times\uv)\cdot\pvp]\tv\cdot [ - g_A^e\, \SP +g_V^e\, \HM ]\}$ \\
${\rm Im} (g_T PF_2^r)$ & $4 E[ - g_A^e\,\SP +g_V^e \,\HM ] \times \Kv\cdot \rv$ \\
${\rm Re} (g_T F_1^{rut})$ &
$ 4 E \left(E[\, r^0 (\uv \times \tvt) - u^0(\rv\times \tvt)
        ]\cdot [-g_V^e\, \SP + g_A^e \HM ]\right. $  \\
& $\left.+ \{[(\rv\times\uv)\cdot\pvp]\tv\cdot [-g_V^e\, \SM + g_A^e\,
\HP ]\}\right)$ \\
${\rm Re} (g_T F_2^r)$ & $4 E [-g_V^e\,\SM+ g_A^e \,\HP ]\times \Kv\cdot \rv$ \\
${\rm Re} (g_T PF_1^{rut})$ & $-4 E^2 \{ 
[(\rvt\cdot\tvt)\uv - (\uvt\cdot \tvt)\rv]  
\cdot 
[g_V^e\,\SP - g_A^e \, \HM]
$ \\
& $ + 
(r^0 u^3-r^3 u^0) \tv 
\cdot 
[g_V^e\,\SM - g_A^e \, \HP ]
\}$ \\
${\rm Re} (g_T PF_2^r)$ &  $4 E^2 \rv\cdot [g_V^e\,\SP - g_A^e \, \HM ]$ \\ \hline
\end{tabular}
\caption{List of $T$ correlations}
\end{center}
\end{table}

\comment{ % gAe table commented out
\begin{table}\label{T_gAe_table}
\begin{center}
\begin{tabular}{||c|c||}\hline
Term & Correlation                         \\ \hline \hline
${\rm Im} g_T F_1^{rut}$ & $4 E^2 \{ \SM \cdot [(\rvt\cdot\tvt)\uv -
                                           (\uvt\cdot \tvt)\rv]+ $ \\
& $\SP\cdot \tv (r^0 u^3-r^3 u^0) \}$ \\
${\rm Im} g_T F_2^r$ &  $-4 E^2 \SM\cdot \rv$ \\
${\rm Im} g_T PF_1^{rut} $ &
$ 4 E^2 [ r^0 (\uv \times \tvt) - u^0(\rv\times \tvt)
        ]\SM $  \\
& $-4 E\{[(\rv\times\uv)\cdot\pvp][\tv\cdot \SP]\}$ \\
${\rm Im} g_T PF_2^r$ & $-4 E [\SP\times \Kv]\cdot \rv$ \\
${\rm Re} g_T F_1^{rut}$ &
$ 4 E^2 [ r^0 (\uv \times \tvt) - u^0(\rv\times \tvt)
        ]\HM +$  \\
& $4 E\{[(\rv\times\uv)\cdot\pvp][\tv\cdot \HP]\}$ \\
${\rm Re} g_T F_2^r$ & $4 E [\HP\times \Kv]\cdot \rv$ \\
${\rm Re} g_T PF_1^{rut}$ & $-4 E^2 \{ \HM \cdot [(\rvt\cdot\tvt)\uv -
                                           (\uvt\cdot \tvt)\rv]+ $ \\
& $\HP\cdot \tv (r^0 u^3-r^3 u^0) \}$ \\
${\rm Re} g_T PF_2^r$ &  $-4 E^2 \HM\cdot \rv$ \\ \hline
\end{tabular}
\caption{List of $T$ correlations for $g_{A}^e$}
\end{center}
\end{table}
}

\section{CP and T properties of correlations}\label{characterization}

It is important to characterize the C, P and T properties of the various
terms in the correlations, which would in turn depend on the
corresponding properties of the structure functions which occur in them. 

In this context we recall that a similar analysis was done for  the
one-particle inclusive case treated in \cite{BASDR}. In that case, we
 deduced the important result that when the final state consists of
a particle and its anti-particle, it is not possible to have any CP-odd
term in case of $V$ and $A$ BSM interactions. This deduction depended
on the property that in the centre-of-mass frame, the particle and
anti-particle three-momenta are equal and opposite. In the present case
of two-particle inclusive distributions, even if the two particles
observed are conjugates of each other, their momenta are not
constrained. Thus it is possible to have CP-odd correlations even in the
$V,A$ case.

We now come to a more systematic analysis. We consider two important
cases, one when the particles $h_1$ and $h_2$
in the final state in $e^+e^-\to h_1h_2 X$ are their own conjugates, and
the other when they are not.
We treat these two cases separately.

\subsection{Case A: $h_1^c = h_1$, $h_2^c = h_2$}

In this case, the first entry in Table 1 is CP even for $\rv\equiv \pv_1,
\pv_2$, that is, when $\rv$ is an ordinary (polar) vector,
and CP odd for $\rv\equiv \nv$, a pseudo-vector (axial vector). 
The same is true for the fourth entry in
Table 1. On the other hand, the second and third entries are CP odd for 
$\rv\equiv \pv_1, \pv_2$ and CP even for $\rv\equiv \nv$. Moreover, it can be
checked that the first two entries are CPT even, where T is na\" ive
time reversal, and the last two entries are CPT odd. (Henceforth,
whenever we refer to T, we will mean na\" ive time reversal). 
This implies, via
the CPT theorem, that the last two entries need the  absorptive part
to be nonzero.\footnote{For a review, see \cite{Rindani:1994ad}.
 Normally, such an absorptive part would 
simply be indicated
by the occurrence of the imaginary part, rather than the real part, of
the relevant structure function. However, the definition of $F^r$ in eq.
(\ref{scalarH}) needs an extra factor of $i$ for this to happen. Such an
$i$ would be natural if the $F^r$ were defined to be real for point
couplings. However, we have maintained the definitions of ref.
\cite{DR2}.} 

In Table 2, the entries with $W_1$ and those with $W_4$ are
CP even, the former being even under na\" ive T and the latter odd.
Of the other entries, the terms corresponding to $W_2^{rt}$ are CP even
when both $\rv$ and $\tv$ are ordinary vectors, and CP odd when
one of $\rv$ and $\tv$ is an ordinary vector and the other is a pseudo-vector. 
Of these four
rows in Table 2, the first two are T even and the next two T odd for
the cases when these are all CP even, and the opposite T properties hold
when they are all CP odd. The entries with $W_3^{uv}$ are CP odd and T
even for all $u,v$. In this case of $V$ and $A$ couplings, we see that
terms which are even under CPT occur with the real part of the structure
function, whereas those which are odd come with the imaginary part of
the structure function.

In Table 3, the entries corresponding to $F_1^{rut}$ and $F_2^r$ are CP
even when $\uv$, $\tv$ are ordinary vectors (since $\rv$ can only be an
ordinary vector), but CP odd when one of
$\uv$, $\tv$ is a pseudo-vector.
On the contrary, the entries corresponding to $PF_1^{rut}$ and $PF_2^r$
are CP
odd when $\uv$, $\tv$ are ordinary vectors and CP even when one of
$\uv$, $\tv$ is a pseudo-vector.
Deductions regarding the presence or absence of absorptive parts follow
on use of the CPT theorem after noting that when all vectors are
ordinary vectors, 
the first two entries are T even, the next four are T odd, and the last
two are again T even.\footnote{As noted in the previous footnote, but
for the unfortunate absence of factors of $i$ in the definition of 
each of the tensor
structure functions in eq. (\ref{tensorG}), the CPT-even entries would
be associated 
with real parts of structure functions, and the CPT-odd entries with
the imaginary parts.}

\subsection{Case B: $h_1^c \neq h_1$, $h_2^c \neq h_2$}
 
In the case when $h_1$ and $h_2$ are not self-conjugate, the above
statements under case A about the CP properties 
would be true for even linear combinations of the structure functions
for production of $h_1$ and $h_2$ with the structure functions for 
the production of the conjugates of
$h_1$ and $h_2$. For the odd linear combinations, the opposite CP
properties would hold. 

A special case worth considering is when $h_1$ and $h_2$ are conjugates
of each other. In that case, one can decompose each term into a part
which is even under interchange of the four-vectors $p_1$ and $p_2$,
keeping in mind that the structure functions are functions of the invariants
$p_1^2$, $p_2^2$, $p_1\cdot p_2$, $q\cdot (p_1+p_2)$, and $q^2$, which
are even under interchange of $p_1$ and $p_2$, and of $q\cdot
(p_1-p_2)$, which is odd under that interchange. 
 One has now to go through the previous analysis done in
the case when the final-state particles were self-conjugate, resulting
in somewhat different results. In general, there would again be
different combinations of structure functions which would contribute CP-even
and CP-odd terms. However, in some cases, the existing terms transform
into themselves, with a factor of $\pm 1$. In this case, the CP property
is the same or opposite to that in the case of self-conjugate final
state. 

\section{The effect of beam polarization}\label{sec:compol}

We now make some general deductions from the tables on the dependence of
distributions on beam polarization.
We include in this discussion the correlations obtained in \cite{BASDR}
for the one-particle inclusive process as well.
To make the discussion in this section self-contained we repeat some
observations already made in Sec. \ref{correlations}.

The first observation that one can make is that the interference of the 
scalar/pseudoscalar
and tensor BSM interactions with the SM contribution cannot be studied 
unless the electron and/or positron beams are polarized. 
Not only that, it is not sufficient to have longitudinal polarization. A
nonzero transverse polarization is needed to observe the interference
terms. This is easy to understand -- in the limit of vanishing electron
mass, the scalar and tensor couplings are chirality violating, whereas
the vector and axial-vector SM couplings are chirality conserving. Thus,
the two do not interfere, even for arbitrary longitudinal polarization.

The interference of the vector and axial-vector BSM contributions with
the SM contributions, on the other hand, is nonzero for unpolarized beams
as well as polarized beams. In the case of transverse polarization, 
it is necessary for both electron and positron beams to be polarized for
a nonzero answer, in contrast to the case of scalar and tensor BSM
interactions, where either electron or positron beam had to be
polarized.

The second observation in the case of vector and axial-vector BSM
interactions is that the structure functions which contribute when polarization
is included are the same as the ones which contribute when beams are
unpolarized, provided absorptive parts are neglected. 
We assume here that the final-state particles which are observed are
themselves eigenstates of CP, in which case, 
the imaginary parts of the structure functions contain
absorptive parts of the BSM amplitudes.
In other words, no qualitatively 
new information is contained in the polarized distributions if we
neglect the imaginary parts of the structure functions.
This observation, which here is for a general $s$-channel BSM process,
was made in the context of the process $e^+e^- \to HZ$ in
\cite{hagiwara} with anomalous $\gamma ZH$ and $ZZH$ vertices, and
confirmed in \cite{RR1} for general $e^+e^-HZ$ contact interactions.
Such an observation was also made in an older context in \cite{icq} for
the process $e^+e^- \to 3\, {\rm jets}$.

This observation is important because most BSM interactions are
chirality conserving in the limit of massless electrons, and can
therefore be cast in the form of vector and axial-vector couplings.
Thus, in a large class of contexts and theories, it is possible to
conclude that polarization does not give qualitatively new information,
unless absorptive parts are involved. 
This argument can be turned
around, and it is possible to conclude that polarization can be used to
get information on absorptive parts of structure functions  of BSM interactions,
which cannot be obtained with only unpolarized beams.

As a caveat, we note the following: It should not be construed that 
polarization does not play any
positive role for chirality-conserving interactions even when there are
no absorptive parts. In various case, it is possible to enhance the
sensitivity to BSM interactions with a judicious choice of signs of the
polarization. Thus even when no new structure functions are uncovered by
polarization, the information on structure functions which can be obtained with
polarized beams can be quantitatively better than that obtained with
unpolarized beams. 

In our case, if absorptive parts are included, there is
a contribution from ${\rm Im}~W_3$ for the one-particle inclusive
case considered in \cite{BASDR}, and ${\rm Im}~W_3^{uv}$ in the
two-particle inclusive case discussed in this paper. Again, in this case, it
possible to predict the differential cross section for the polarized
case, if the unpolarized cross section is known.

On the other hand, we see that ${\rm
Im}~W_2^{rt}$ (${\rm Im}~W_2$ in the one-particle case) contribute only 
for transversely polarized beams. Thus,
to observe these structure functions, it is imperative to have transverse
polarization, at least of one beam. A further point to notice about the
contribution of ${\rm Im}~W_2^{rt}$ is that if $g_V^e = g_V$ and $g_A^e
= g_A$, the contribution vanishes. In other words, if the new physics
contribution corresponds to the exchange of the same gauge boson as the
SM contribution, so that the coupling at the $e^+e^-$ vertex is the
same, even though the final state may be produced through a new vertex,
the contribution to the distribution is zero. Thus, in case of a neutral
final state, where the SM contribution through a virtual photon vanishes
at tree level,  the
observation of ${\rm Im}~W_2^{rt}$ through transverse polarization
could be used to determine the absorptive part of a loop contribution
arising from $\gamma$ exchange. In case of a charged-particle final
state for which both $Z$ and $\gamma$ contribute, such a contribution
would be sensitive to loop effects arising in both these exchanges.

\section{Some specific processes}\label{exclusive}
In this section, we will examine some exclusive processes that
are of importance at ILC energies, which use the possibility of
longitudinal and transverse polarization of either or both the
beams to enhance the sensitivity to physics beyond the SM.
Typically, the latter may be described in processes where
the final state contains SM particles, as model-independent
form factors or higher-dimensional operators.  Of special interest
to us are the processes $e^+ e^- \to H Z$, $e^+ e^-\to f \overline{f}$.
These processes present an opportunity to demonstrate the efficacy of
the framework studied here.  

Since the final state in these processes has two particles, the
processes are primarily described by the one-particle inclusive
formalism of \cite{BASDR}. However, when one or both of these particles decay
they give rise to a minimum of three particles in the final state, for
which our present formalism would be applicable. 

\subsection{$e^+ e^- \to H Z$}

The process $e^+ e^- \to H Z$ is an important mechanism for the
production of the Higgs in SM. There have been suggestions
\cite{hagiwara,gounaris,cao,han,biswal} that
distributions of the $Z$ or those of the decay products of the $Z$ can
be used to probe a Higgs boson in a multi-Higgs model. The process 
has been recently studied as
a possible setting to study BSM interactions arising from a four-point
$e^+e^-HZ$ coupling~\cite{RR1}.
This has also been extended in \cite{RR2} to include leptonic
decays of the $Z$. 
All possible interactions consistent with Lorentz invariance
are written down in terms of completely general four-point
interactions that characterize new physics in this process.
In particular, these interactions are classified into whether
or not they are chirality conserving or chirality violating.
The former contains terms in the effective vertex with an odd number
of Dirac $\gamma$ matrices, while the latter contain an
even number.  The former set is given by $V_i,\, A_i,\, i=1,2,3$ and
the latter set is given by $S_i,\, P_i,\, i=1,2,3$ where
each of these form factors, taken to be independent of
the Mandelstam variables $s,\, t$ in the process, can be complex.
In ref.~\cite{RR1} the contributions of these to the differential
cross section is evaluated.  In ref.~\cite{RR2}, the differential cross
section after the inclusion of $Z$ decay into a pair of leptons
$\ell\bar \ell$, where $\ell$ is different from $e$, is evaluated.
We treat these two cases separately.

Taking up the the process with $Z$ in the final state as discussed in
\cite{RR1}, it may be readily observed that in
this process, all the $V_i,\, A_i$ contribute to the transverse
polarization cross sections.  One may immediately conclude from this
observation that the spin correlations these generate are analogous
to those generated by $W_2$ by inspection of tables in
ref.~\cite{BASDR}.
A further inspection of the tables will reveal that if the real part
of a certain $V_i$ or $A_i$ contributes to the longitudinal cross section,
then the imaginary part will not, and vice versa.  This is borne out
by the explicit expressions given in ref.~\cite{RR1}.  Finally, this
does not preclude the possibility that some of the $V_i$ or $A_i$
will not generate spin-momentum correlations of the type generated by
$W_1$.  In the present case, the study of the explicit results of
ref.~\cite{RR1} reveals that it is only $V_1$ and $A_1$ that also
generate spin-momentum correlations of the type generated by $W_1$.
The $S_i,\, P_i$, on the other hand, as expected, contribute only to
the transversely polarized cross section in accordance with our tables.
Thus we have concretely illustrated how the general formalism that
we have considered here leads to some insights and provides consistency
checks on specific processes.  

The comments made in Sec. \ref{sec:compol} regarding 
the new information about structure functions available from polarization have
to be interpreted carefully in this case. The reason is that the
formalism we use in this paper assumes an $s$-channel exchange of a new
particle. On the other hand, \cite{RR1} deals with contact interactions.
Hence in constructing the BSM tensors in \cite{RR1}, use has been made of
leptonic momenta, which we do not do here. 

In the extension of \cite{RR1} which includes a leptonic decay of $Z$
\cite{RR2}, in addition to the features discussed above, the new feature
is the existence of an additional structure function, viz., $W_4$. This
structure function appears with the vector triple product $\pv_1 \times
\pv_2 \cdot \pv_+$, which is not possible unless two independent momenta
are measured in the final state. This term, for the case of a $H + (Z \to
\ell^+\ell^-)$ in the final state, is CP and T odd, and is nonvanishing
even in the absence of polarization. Such a term would accompany 
CP-violating form factors, viz., $V_3$ and $A_3$ of \cite{RR1,RR2}, and
being CPT even, would be associated with the real parts of these form
factors. The correlation $\pv_1 \times \pv_2 \cdot \pv_+$ with unpolarized
and longitudinally polarized beams has indeed been studied in \cite{RR2}
as  measure of CP violation.

A more direct application of our formalism would be to the case when
only a new $ZZH$
vertex, rather than a contact $e^+e^-HZ$ vertex, is assumed
\cite{gounaris,hagiwara,han,biswal}.
In this case, there are in general three independent couplings, $a_Z$, $b_Z$
and $\tilde b_Z$, and there
exist relations between the structure functions considered in case of contact
interactions \cite{RR1,RR2} and these couplings.
The observation made earlier, regarding the vanishing of the
contribution of ${\rm Im}~W_2^{rt}$ when only the SM gauge boson is
exchanged in new-physics process, now implies that if only a new $ZZH$
vertex, rather than contact interaction is assumed, there will be no
contribution corresponding to the ${\rm Im}~W_2^{rt}$ term.
Alternatively, such a term will be nonvanishing for a $\gamma ZH$
vertex, and the observation of the corresponding transverse polarization
dependence would signal such a $\gamma ZH$ vertex with an
absorptive part.

\subsection{$e^+ e^- \to f \overline{f}$}

In this section we study the process 
$e^+ e^- \to f \overline{f}$, where $f$ is a quark or a lepton,
a process which will dominate at the ILC.
We also look at further decays of the final-state fermions when they are
heavy and
the momentum correlations amongst these as  probes of BSM
interactions.  
We concentrate on CP-odd correlations which indicate CP violation and
are therefore important to study. However, these are by no means the only
interesting correlations. CP-even correlations could be used to study
new CP-conserving interactions like magnetic dipole moments.

Here one may recall the 
one-particle inclusive case discussed in \cite{BASDR}, where it was found that
for the specific process where the final-state is a two-particle state
consisting of a charged particle and its conjugate, 
there can be no CP-violating observables for new physics having a  $V,A$
structure in the absence of polarization. Even in the presence of
transverse polarization, CP violation can be observed only in the case
of scalar/pseudoscalar and tensor interactions \cite{BASDRtt}. However,
CP-odd observables can be constructed if the spin of one of the final
conjugate pair of particles is observed. In the case where one of the
decay products is observed, the polarization of the decaying particle is
being made use of indirectly.
Thus it is expected that
in our analysis of a two-particle inclusive states, it would be possible
to incorporate CP-violating correlations.

Some years ago Hoogeveen and Stodolsky~\cite{HS} considered the
possibility of the observation of the electric dipole moment (edm) of
the electron in high-energy $e^+ e^-$ reactions with transversely
polarized beams.  Although there are stringent bounds on the magnitude
of this observable today, it is interesting to recall the consequences
of their very general arguments.  It was pointed out that the existence
of a transverse polarization vector could be used to construct two
CP-odd variables, that involve (a) a scalar product with the
three momentum of the particle $f$, and (b) a triple product involving
these two momenta and the beam direction.  An inspection of our
tables immediately reveals that such CP-odd spin-momentum correlations
occur in the presence of the form factor $g_T\, PF_1$.

On the other hand, as the experimental constraints on heavier
SM fermion edm's such as the $\tau$ lepton and the top quark are less
stringent, there has been considerable work in trying to probe these
quantities with theorists proposing several tests.
Early work in this regard was the study by
Couture~\cite{PHLTA.B272.404}.
Since these particles
are highly unstable, they decay very rapidly, the $\tau$ often into
a $\rho$ or a $\pi$ along with neutrino emission, while the
top decays into a $b W$.  Therefore, what one considers in
practice is the possibility of probing the edm via momentum
correlations of the decay products, which in reality probes the
spin correlations of the fermion pair $f\bar f$ produced in the reaction, as
the decay is due to the weak interaction which serves as a spin analyzer.
The subject was studied in detail in ref.~\cite{PHRVA.D48.78}
where tensor correlations constructed from
the momenta of the decay products were used as a probe, following
the work done in the context of the weak dipole moment, the analogue
of the edm when the photon is replaced by the $Z$ boson,
see ref.~\cite{ZEPYA.C43.117}.
That the sensitivity of vector correlations to weak-dipole
edm's is enhanced in the presence of longitudinal polarization
was shown in ref.~\cite{PRLTA.73.1215,PHRVA.D51.5996}.

Here we note that the edm of $f$, the fermion being pair produced,
involves a coupling to the initial-state $e^+e^-$ which is
vector or axial-vector in nature, 
in contrast to the edm of the electron which involves a tensor coupling.
Restricting to the case of the final state being produced through $V$,
$A$ interactions with the initial-state $e^+e^-$, 
it is possible to find terms in Table 2, using
suitable choices of vectors $r$, $t$, $u$ and $v$, which correspond to
CP-violating observables. Thus, for example, choosing the two observed
particles to be conjugates of each other, 
it is possible to generate T-odd terms
associated with $W_2^{rt}$, where one of $r$, $t$ is $n$, and the other
is $p_1$ or $p_2$, a combination of which would also be CP odd. The
observable associated with $W_4$ is already CP and T odd. A combination
of $W_3^{uv}$ for suitable choice of $u$ and $v$ can easily be made CP
odd (though even under T). Thus, a number of CP-odd terms can be
generated. In the case of the process under consideration, restricting
to $\gamma$ and $Z$ exchanges alone, these would arise from the edm and
weak dipole moment of $f$.

For unpolarized beams, an explicit expression 
for the differential cross section 
is provided for the case of $e^+ e^-\to \tau^+ \tau^-
\to a(p_1) \overline{b}(p_2) X$ in ref.~\cite{PHRVA.D48.78}, see eq.~(5.2)
therein, for the contribution of the edm to the differential cross-section.
What is of relevance for us here is that this expression contains
scalar products of the beam direction with the sum and
the cross-product of the three momenta of $a,\, \overline{b}$, as well
as a product of each of these with the scalar product of the
beam direction with the difference of the three momenta.
Each of these quantities is generated from our tables as just described.
  
The generalization to
include longitudinal polarization, derived in
\cite{PRLTA.73.1215,PHRVA.D51.5996}, where it was emphasized that CP-odd vector
correlations, which are suppressed in the unpolarized case, are
enhanced with the use of longitudinal beam polarization. This feature
can be seen from our tables, where the vector correlations are
associated with the $W_3$ and $W_4$ terms. In these terms, the
unpolarized correlation is necessarily proportional to $g_V$ or $g_V^e$,
which is numerically small. With longitudinal polarization, the
polarization dependent terms involve $g_A$ or $g_A^e$, providing an
enhancement. For a review for these effects at
high energy $e^+ e^-$ colliders, see ref.~\cite{ARS} and references therein.

The work on CP-odd correlations with longitudinal beam polarization in
$\tau$-pair production \cite{PRLTA.73.1215,PHRVA.D51.5996} 
was subsequently extended to the process $e^+ e^- \to
t \overline{t} \to b W^+ \overline{b} W^-$~\cite{PHLTA.B343.333},
where analogous correlations were studied.
The forms of angular distributions derived in further work on the study of 
dipole moments through charged-lepton \cite{poulose,sdr,grzad} and $b$-quark 
\cite{grzad,christova} single-particle
distributions could also be deduced through the tables of our earlier
work \cite{BASDR}.
Other work where other CP-odd spin and momentum correlations have
been studied can be found in \cite{soniPhysRep}.

This example has some common features with the example of $e^+e^-\to
t\overline t$ through contact interactions discussed in \cite{BASDRtt}
and through leptoquark exchange, discussed in \cite{PHLTA.B602.97}.

\section{Conclusions and Discussion}\label{conclusions}

To recapitulate, we have computed two-particle angular distributions
in $e^+e^-$ collisions
arising from the interference between the virtual $\gamma$ and $Z$
exchange SM amplitudes and BSM amplitudes characterized by their Lorentz
signatures, with the unknown physics lumped into structure functions.
Transverse and longitudinal beam polarizations are explicitly included.
We have presented a discussion on the nature of the correlations and the 
deductions that can be made on their polarization dependence.
We have also discussed the CP and CPT properties of certain structure
functions. We then specialized to specific final states $HZ$ and $f\bar
f$. In case of the $HZ$ final state we find some subtle effects that are
absent when only one-particle inclusive processes are considered. A
summary of a variety of effects due to popular sources of BSM physics as
manifested in the present framework is provided. 
Our work demonstrates the power of the general model-independent
framework and justifies the extension of results known for one-particle
distributions to two-particle distributions. 

As discussed in Sec. \ref{intro}, processes which involve $t-$ and 
$u-$channel contributions either at the
level of SM or in the new physics, strictly speaking, lie outside the
scope of the present formalism. In those cases, not only do the
structure functions we use depend on additional invariants involving the
electron and positron momenta, even the tensors written down for the
interference of the SM terms with the BSM terms would involve the
electron and positron momenta. This requires further study.
Such a formalism when developed would find application to several
processes where $t-$ or $u-$channel exchanges could contribute to the SM
amplitude, as for example in $W^+W^-$ production, wherein
transverse-polarization effects were studied in \cite{PHRVA.D49.2174}, 
or in $\gamma Z$
production, polarization effects in which were studied in
\cite{BASDRgammaZ}. An
extension of such a formalism would be useful even when considering
interference effects between $s-$channel and $t-$channel exchanges in a
final state like a pair of neutralinos in the supersymmetric extension
of SM. Such a process was studied for example in \cite{choi}. Another
situation where an extended formalism would be useful is when the new
physics is expressed as four-point contact interactions, as for example
in \cite{basdrcont,RR1,RR2}

Another class of interesting processes involve BSM spectrum of
particles, as for example in a theory like the minimal supersymmetric
standard model (MSSM). A description of such processes would also
involve a modification of our formalism, not only to include new
particles, but also to include $t$- and $u$-channel exchanges.

Some popular scenarios, such as extra-dimensional
models, non-commutative models, contact interactions, etc., 
could also be studied with a more
general formalism indicated. One would expect to make some general
predictions in these cases.

Even though Dass and Ross in their paper \cite{DR2} discuss the
one-particle inclusive process where the spin of the observed particle
is also measured, we have not treated this topic in our context in this
work. This would be an interesting future study. In one of the examples we
consider, the final state arises from the decay of a real or virtual
spin-1 state. Hence the analysis presented in \cite{DR2}, which was for
a decaying spin-$\frac{1}{2}$ particle, is not directly applicable.
It would be useful to extend the formalism of \cite{DR2} to spin 1.

\bigskip

\noindent {\bf Acknowledgements:}
BA thanks the Council for Scientific and
Industrial Research for support during the course of these
investigations under scheme number 03(0994)/04/EMR-II, as well
as the Department of Science and Technology, Government of India.
SDR acknowledges partial support from the IFCPAR project no. 3004-2.

\end{document}